\documentclass{desyproc}

\begin{document}
\title{Solution to the Isotropy Problem for\\ Cosmological Hidden Vector Models}

\author{{\slshape J.A.R. Cembranos, A.L. Maroto and S.J. N\'u\~nez Jare\~no}\\[1ex]
Departamento de  F\'{\i}sica Te\'orica I, Universidad Complutense de Madrid, E-28040 Madrid, Spain}

\contribID{cembranos\_jose}

\desyproc{DESY-PROC-2013-XX}
\acronym{Patras 2013} 
\doi  

\maketitle

\begin{abstract}
Gauge bosons associated to new gauge symmetries under which the standard model particles are not charged are predicted in many extensions of the standard model of particles and interactions. We show that under very general conditions, the average energy-momentum tensor of these rapidly oscillating vector
fields is isotropic for any locally inertial observer. This result has a fundamental importance in order to consider coherent vector fields as a viable alternative to support models of dark matter, dark energy or inflation.
\end{abstract}

\section{Introduction}

Despite the large improvement in our knowledge about cosmology in the last decades, there are various puzzles associated with basic features of the history of the universe. For instance, observational data favour the existence of inflation in the early universe, or the existence of dark matter and dark energy at later times. However, the intimate nature of these components remains undetermined. A possible solution to these questions has been formulated in terms of coherent
rapid oscillations of bosonic fields. In this context, scalar models have been traditionally proposed for inflation \cite{Damour}. Scalar massive particles, such as axions \cite{axions}, or other massive scalar \cite{scalars} or pseudoscalar particles \cite{branons} have been considered standard candidates as non-thermal relics. Oscillating scalar fields have also been studied as dark energy models \cite{Liddle}.

\section{Hidden vector models in cosmology}

The same approach to these open questions is offered by vector fields. Indeed, a large number of vector models have been studied in relation
with cosmology \cite{Vectors}. For instance, inflationary models can be supported by vectors \cite{Ford,Golovnev,Gflation,Soda,vectorinflation,Peloso}.
Isotropic and homogeneous triad configurations of non-abelian vector gauge bosons have been recently considered as a viable model that can be
supported even by the standard QCD action \cite{Chromoinflation} (read \cite{marco} however).

Vector modes have been also studied as the origin of metric perturbations in the so called curvaton scenario \cite{Dimopoulos}.  Coherent oscillations of massive vector fields have been analyzed as non-thermal dark matter candidates in \cite{Nelson} and its phenomenology merges within general hidden photons
models \cite{HP1,HP2,HP3}.
There are also a rich variety of vector dark energy models, with potential terms \cite{DE} or without them \cite{VT}.

\begin{figure}
\centerline{\includegraphics[width=1.0\textwidth]{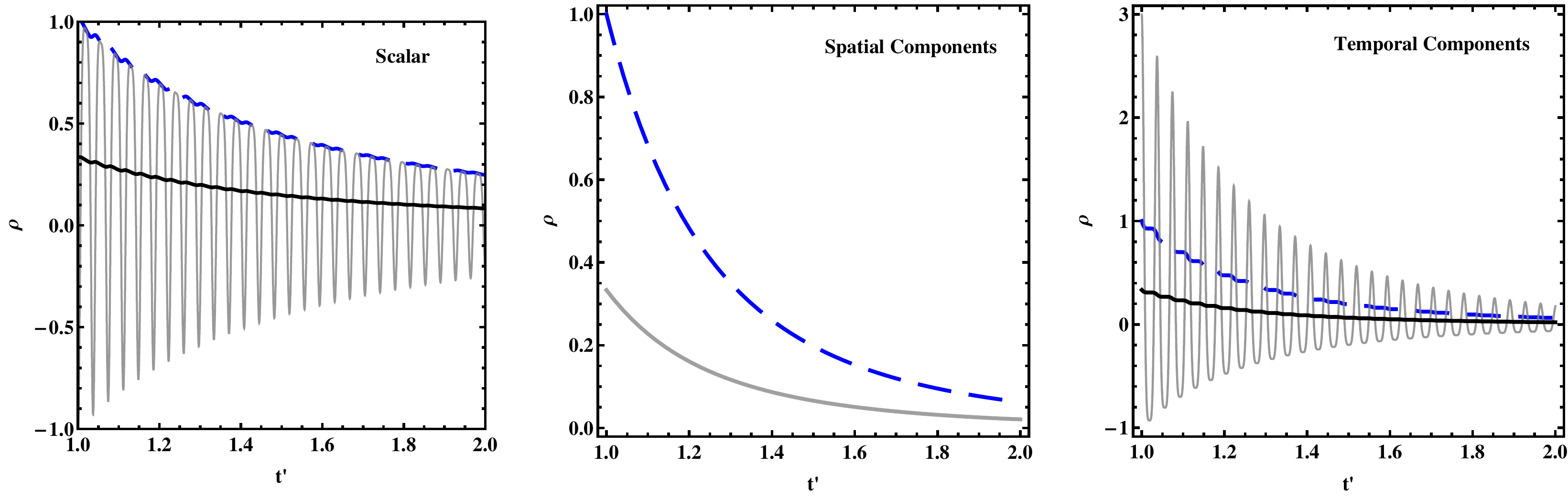}}
\caption{Graphical representation of different energy components of a scalar field (left), spatial components of a vector field (center) and its temporal component (right) for a quartic potential. In particular, one can observe the evolution of the average pressure along the three spatial directions $p$ (grey line), energy density (blue dashed line) and average pressure $\langle p \rangle = \langle \rho \rangle/3$ (black line). The evolution is compute in a radiation dominated-universe, the y-axis is normalized to the initial value $\rho(t'_0)$ and time, $t'$, is in $H_0^{-1}$ units. For this concrete case,
the spatial components of a vector field verify $\langle p \rangle = p$ \cite{Isotropy}.}\label{Fig:p}
\end{figure}

\section{The isotropy problem}

However, there is a generic problem associated to cosmological vector models. The dynamics of a homogenous vector field is necessarily anisotropic
and these models are generically excluded by anisotropy constraints imposed by CMB observations.
In any case, there have been several proposals in the literature to solve this isotropy problem.
As we have already commented in the previous section, the presence of several vector modes can compensate their intrinsic anisotropy
by tuning their global structure. In particular, triad configurations corresponding to SU(2) gauge groups have been extensively studied
\cite{solution, Galtsov, Zhang,Armendariz,Chromoinflation}.

A simpler solution arises if only the temporal components of the vector fields are allowed to evolve \cite{VT}. On the other hand, the isotropy violation of vector models can be alleviated by increasing the number of vector modes. A large number $N$ of randomly oriented vector fields reduce the amount of anisotropy  by a factor $1/\sqrt{N}$  \cite{Golovnev}.

\section{Averaged isotropy of rapid evolutions}

Another possibility was analyzed in Ref. \cite{Dimopoulos}, where it was shown that a homogenous linear polarized Abelian coherent vector
mode, oscillating with a quadratic potential, has associated an isotropic averaged energy-momentum tensor.

This idea has been recently extended to any kind of polarization and potential by means of the {\it isotropy theorem} \cite{Isotropy}. In fact, it has also been shown to be independent of the Abelian character of the vector field. The theorem guarantees the isotropy of the averaged energy-momentum tensor for any type of initial configuration provided that the vector evolution is bounded and rapid compared to the metric evolution. A paradigmatic case is provided by massive vector fields with  masses larger than the Hubble parameter in a Robertson-Walker geometry, but there is a large number of possibilities (see Fig. \ref{Fig:p} for a different example).

\section{Conclusions}

Coherent homogeneous vector fields can be the origin of the unidentified cosmological constituents,
such as the inflaton, the dark matter or the dark energy. This possibility is thought to suffer
important constraints on anisotropies imposed by different astrophysical observations.
However, a general isotropy theorem for vector fields have proved that this is not the case for
models based on bounded rapid vector evolutions.

\section*{Acknowledgments}

This work has been supported by MICINN (Spain) project numbers FIS 2008-01323, FIS2011-23000, FPA2011-27853-01 and Consolider-Ingenio MULTIDARK CSD2009-00064.


\begin{footnotesize}

\end{footnotesize}



\begin{thebibliography}{99}
%
\bibitem{Damour}
  T.~Damour and V.~F.~Mukhanov,
  Phys.\ Rev.\ Lett.\  {\bf 80} (1998) 3440;
 A.~R.~Liddle and A.~Mazumdar,
  Phys.\ Rev.\ D {\bf 58} (1998) 083508.
%
\bibitem{axions}
R.D. Peccei, H.R. Quinn, Phys. Rev. Lett. \textbf{38} (1977) 1440; 
Phys. Rev. \textbf{D16} (1977) 1791 ;  
S.~Weinberg, Phys. Rev. Lett. \textbf{40} (1978) 223; 
F.~Wilczek, Phys. Rev. Lett. \textbf{40} (1978) 279. 
%
\bibitem{scalars}
  B.~de Carlos, J.~A.~Casas, F.~Quevedo and E.~Roulet, Phys.\ Lett.\ B {\bf 318}, 447 (1993); 
  J.~A.~R.~Cembranos, Phys.\ Rev.\ Lett.\  {\bf 102}, 141301 (2009); 
  J.\ Phys.\ Conf.\ Ser.\  {\bf 315}, 012004 (2011). 
%
\bibitem{branons}
  A. Dobado and A. L. Maroto, Nucl. Phys. B \textbf{592}, 203 (2001);
  J.~Alcaraz {\it et al.}, Phys. Rev.{\bf D67}, 075010 (2003); 
  J.~A.~R.~Cembranos, A.~Dobado and A.~L.~Maroto, Phys. Rev. {\bf D65} 026005 (2002); 
  Phys.\ Rev.\ Lett.\  {\bf 90}, 241301 (2003); 
  A.~L.~Maroto, Phys.\ Rev.\ D {\bf 69}, 043509 (2004); 
  Phys.\ Rev.\ D {\bf 69}, 101304 (2004). 
%
\bibitem{Liddle}  A.~R.~Liddle and R.~J.~Scherrer,
  Phys.\ Rev.\ D {\bf 59} (1999) 023509;
 S.~Dutta and R.~J.~Scherrer,
  Phys.\ Rev.\ D {\bf 78}, 083512 (2008).
%
\bibitem{Vectors}
  JCAP {\bf 0902}, 025 (2009); 
  J.~Beltran Jimenez, R.~Lazkoz and A.~L.~Maroto,
  Phys.\ Rev.\ D {\bf 80}, 023004 (2009); 
  J.~Beltran Jimenez and A.~L.~Maroto,  
  Phys.\ Rev.\ D {\bf 80}, 063512 (2009);
  Int.\ J.\ Mod.\ Phys.\ D {\bf 18}, 2243 (2009); 
  J.~Beltran Jimenez, T.~S.~Koivisto, A.~L.~Maroto and D.~F.~Mota,  
  JCAP {\bf 0910}, 029 (2009); 
  J.~Beltran Jimenez and A.~L.~Maroto,  
  JCAP {\bf 1012}, 025 (2010);
  Phys.\ Rev.\ D {\bf 83}, 023514 (2011);
  J.~Beltran Jimenez and A.~L.~Maroto,  
  Mod.\ Phys.\ Lett.\ A {\bf 26}, 3025 (2011);
  E.~Carlesi, A.~Knebe, G.~Yepes, S.~Gottloeber, J.~Beltran Jimenez and A.~L.~Maroto,
  arXiv:1205.1695 [astro-ph.CO];  
  J.~Beltran Jimenez, A.~L.~Delvas Froes and D.~F.~Mota,
  Phys.\ Lett.\ B {\bf 725}, 212 (2013);  
  J.~Beltran Jimenez, R.~Durrer, L.~Heisenberg and M.~Thorsrud,
  arXiv:1308.1867 [hep-th].  
%
\bibitem{Ford} L.~H.~Ford,
  Phys.\ Rev.\ D {\bf 40} (1989) 967.
%
\bibitem{Golovnev}
      A.~Golovnev, V.~Mukhanov and V.~Vanchurin,      
      JCAP {\bf 0806} (2008) 009.
%
\bibitem{Gflation}
A.~Maleknejad, M.~M.~Sheikh-Jabbari and J.~Soda,  
  Phys.\ Rept.\  {\bf 528}, 161 (2013).
\bibitem{Soda}
  K.~Yamamoto, M.~-a.~Watanabe and J.~Soda,
  Class.\ Quant.\ Grav.\  {\bf 29}, 145008 (2012);
  M.~-a.~Watanabe, S.~Kanno and J.~Soda,
  Phys.\ Rev.\ Lett.\  {\bf 102}, 191302 (2009);
  K.~Murata and J.~Soda,
  JCAP {\bf 1106}, 037 (2011);
  A. Maleknejad, M. M. Sheikh-Jabbari,
  Phys. Rev. D 85, 123508 (2012).
%
\bibitem{vectorinflation}
  T.~Koivisto and D.~F.~Mota, JCAP {\bf 0808}, 021 (2008); 
  K.~Bamba, S.~'i.~Nojiri and S.~D.~Odintsov, Phys.\ Rev.\ D {\bf 77}, 123532 (2008). 
%
\bibitem{Peloso}
  B.~Himmetoglu, C.~R.~Contaldi and M.~Peloso, Phys.\ Rev.\ Lett.\  {\bf 102}, 111301 (2009);
  A.~E.~Gumrukcuoglu, B.~Himmetoglu and M.~Peloso, Phys.\ Rev.\ D {\bf 81}, 063528 (2010). 
%
\bibitem{Chromoinflation}
  A.~Maleknejad and M.~M.~Sheikh-Jabbari, Phys.\ Lett.\ B {\bf 723}, 224 (2013);
  Phys.\ Rev.\ D {\bf 84}, 043515 (2011); 
  P.~Adshead and M.~Wyman,
  Phys.\ Rev.\ Lett.\  {\bf 108}, 261302 (2012); 
  Phys.\ Rev.\ D {\bf 86}, 043530 (2012); 
  K.~Yamamoto,
  Phys.\ Rev.\ D {\bf 85}, 123504 (2012); 
  M.~M.~Sheikh-Jabbari,
  Phys.\ Lett.\ B {\bf 717}, 6 (2012); 
  A.~Ghalee,
  Phys.\ Lett.\ B {\bf 717}, 307 (2012); 
  M.~Noorbala and M.~M.~Sheikh-Jabbari,
  arXiv:1208.2807 [hep-ph]; 
  K.~-i.~Maeda and K.~Yamamoto,
  arXiv:1210.4054 [astro-ph.CO];  
  E.~Dimastrogiovanni, M.~Fasiello and A.~J.~Tolley,
  arXiv:1211.1396 [hep-th].  
%
\bibitem{marco}
  E.~Dimastrogiovanni and M.~Peloso,  Phys.\ Rev.\ D {\bf 87}, 103501 (2013);
  R.~Namba, E.~Dimastrogiovanni and M.~Peloso,  arXiv:1308.1366 [astro-ph.CO].
%
\bibitem{Dimopoulos}
   K.~Dimopoulos,
    Phys.\ Rev.\ D {\bf 74} (2006) 083502.
%
\bibitem{Nelson}
    A.~E.~Nelson and J.~Scholtz,
    Phys.\ Rev.\ D {\bf 84} (2011) 103501.
%
\bibitem{HP1}
  Raffelt GG., {\it Phys. Rev. D} 33, 897 (1986); 
  Raffelt GG, Dearborn DSP., {\it Phys. Rev. D} 36, 2211 (1987); 
  V. Popov and O.Vasil'ev, Europhys. Lett.  {\bf15}, 7 (1991);
  V. Popov, Turkish Journal of Physics  {\bf23}, 943 (1999);
  Schlattl H, Weiss A, Raffelt GG., {\it Astropart. Phys.} 10, 353 (1999);
  Gondolo P, Raffelt GG., arXiv:0807.2926 [astro-ph] 
  Raffelt GG., {\it Lect.  Notes Phys.} 741, 51 (2008);
  Redondo J., {\it JCAP} 0807:008 (2008);
  Redondo J, Postma M.  {\it JCAP} 0902:005 (2009);
  Jaeckel J, et al.,  {\it Phys.\ Rev.\  D} 75:013004 (2007);
  M. Pospelov, A. Ritz, and M.B. Voloshin,  Phys. Lett. B {\bf 662}, 53 (2008);
  B. Batell, M. Pospelov, and  A. Ritz,  Phys. Rev. D {\bf 80}, 095024 (2009);
  H.~An, M.~Pospelov and J.~Pradler, Phys.\  Rev.\  Lett.\  111, {\bf 041302} (2013).
%
\bibitem{HP2}
  S.~A.~Abel, M.~D.~Goodsell, J.~Jaeckel, V.~V.~Khoze, A.~Ringwald,
  JHEP {0807}, 124 (2008); 
  JHEP {0911}, 027 (2009); 
  M.~Cicoli, M.~Goodsell, J.~Jaeckel, A.~Ringwald,
  J.~Jaeckel and A.~Ringwald, Ann.\ Rev.\ Nucl.\ Part.\ Sci.\  {\bf 60}, 405 (2010);
  M.~Goodsell, A.~Ringwald, Fortsch.\ Phys.\  {58}, 716 (2010);
  J. Jaeckel and A. Ringwald, Ann. Rev. Nucl. Part. Sci. {\bf 60}, 405 (2010);
  M.~Goodsell, J.~Jaeckel, J.~Redondo, A.~Ringwald,
  JHEP {1107}, 114 (2011); 
  P.~Arias, D.~Cadamuro, M.~Goodsell, J.~Jaeckel, J.~Redondo, A.~Ringwald,
  JCAP {1206}, 013 (2012); 
  M.~Goodsell, S.~Ramos-Sanchez, A.~Ringwald,
  JHEP {1201}, 021 (2012); 
  A.~Ringwald,  Phys.\ Dark Univ.\  {\bf 1}, 116 (2012).
%
\bibitem{HP3}
  K.~Baker, G.~Cantatore, S.~A.~Cetin, M.~Davenport, K.~Desch, B.~Döbrich, H.~Gies and I.~G.~Irastorza {\it et al.},
  Annalen Phys.\  {\bf 525}, A93 (2013); 
  J.~Jaeckel and J.~Redondo,    
  arXiv:1307.7181 [hep-ph]; 
  arXiv:1308.1103 [hep-ph];  
  D.~Horns, J.~Jaeckel, A.~Lindner, A.~Lobanov, J.~Redondo and A.~Ringwald,
  JCAP {\bf 1304}, 016 (2013); 
  J.~Redondo and G.~Raffelt,  
  JCAP {\bf 1308}, 034 (2013).
%
\bibitem{DE}
 C.~G.~Boehmer and T.~Harko,
  Eur.\ Phys.\ J.\ C {\bf 50} (2007) 423.
%
\bibitem{VT} J.~Beltran Jimenez and A.~L.~Maroto,
  Phys.\ Rev.\ D {\bf 78} (2008) 063005;
  J.~Beltran Jimenez and A.~L.~Maroto,
  JCAP {\bf 0903} (2009) 016;
  J.~Beltran Jimenez and A.~L.~Maroto,
  Phys.\ Lett.\ B {\bf 686} (2010) 175;
  E.~Carlesi, A.~Knebe, G.~Yepes, S.~Gottloeber, J.~Beltran Jimenez and A.~L.~Maroto,
  MNRAS {\bf 418} (2011) 2715, arXiv:1108.4173 [astro-ph.CO].
%
\bibitem{solution}
  J. Cervero and L. Jacobs,
  Phys. Lett. B 78, 427 (1978);
  M. Henneaux,
   J. Math. Phys. 23, 830 (1982);
  Y. Hosotani,
  Phys. Lett. B 147, 44 (1984).
%
\bibitem{Galtsov}
  D. V. Galtsov and M. S. Volkov,
  Phys. Lett. B 256, 17 (1991)
  D.~V.~Gal'tsov,
  arXiv:0901.0115 [gr-qc].
%
\bibitem{Zhang} Y. Zhang, Phys. Lett. B340 (1994) 18; Class. Quan. Grav. 13 (1996) 2145;
 E.~Elizalde, A.~J.~Lopez-Revelles, S.~D.~Odintsov and S.~Y.~Vernov,
    arXiv:1201.4302 [hep-th].
%
\bibitem{Armendariz} C.~Armendariz-Picon,
        JCAP {\bf 0407} (2004) 007.
%
\bibitem{Isotropy}
  J.~A.~R.~Cembranos, C.~Hallabrin, A.~L.~Maroto and S.~J.~N.~Jareno,
  Phys.\ Rev.\ D {\bf 86}, 021301 (2012); 
  J.~A.~R.~Cembranos, A.~L.~Maroto and S.~J.~N.~Jareno,
  Phys.\  Rev.\  D 87, {\bf 043523} (2013). 

\end{thebibliography}
\end{document}